\documentclass[twocolumn,prb,amsmath,amssymb,aps,superscriptaddress,floatfix]{revtex4-2}
\usepackage{CJK}
\usepackage{graphicx}
\usepackage{color}
\usepackage{braket}
\usepackage{upgreek}
\usepackage{xr-hyper}
\makeatletter
\newcommand*{\addFileDependency}[1]{% argument=file name and extension
  \typeout{(#1)}
  \@addtofilelist{#1}
  \IfFileExists{#1}{}{\typeout{No file #1.}}
}
\makeatother

\newcommand*{\myexternaldocument}[1]{%
    \externaldocument{#1}%
    \addFileDependency{#1.tex}%
    \addFileDependency{#1.aux}%
}

\myexternaldocument{suppl}
\usepackage{dcolumn} 
\usepackage{bm}     
\usepackage{amsfonts}
\usepackage[bookmarks=false,colorlinks,citecolor=blue]{hyperref}
\hypersetup{colorlinks=true,linkcolor=blue,filecolor=blue,citecolor = blue, urlcolor=blue}\usepackage{amsmath}

\usepackage[version=4]{mhchem}
\usepackage{siunitx}

\newcommand{\bfl}{\begin{flushleft}}
\newcommand{\efl}{\end{flushleft}}

\usepackage{multirow}
\usepackage{array}
\newcolumntype{P}[1]{>{\centering\arraybackslash}p{#1}}

\usepackage{subcaption}   
\usepackage{caption}
\begin{document}

\title{Direct Comparison of Magnetic Penetration Depth in Kagome Superconductors AV$_3$Sb$_5$ (A = Cs, K, Rb)}

\date{\today}

\author{Austin R. Kaczmarek}
\affiliation{Laboratory of Atomic and Solid-State Physics, Cornell University, Ithaca, NY, USA}

\author{Andrea Capa Salinas}
\affiliation{Materials Department, University of California Santa Barbara, Santa Barbara, CA , USA}

\author{Stephen D. Wilson}
\affiliation{Materials Department, University of California Santa Barbara, Santa Barbara, CA , USA}

\author{Katja C. Nowack}
\affiliation{Laboratory of Atomic and Solid-State Physics, Cornell University, Ithaca, NY, USA}
\affiliation{Kavli Institute at Cornell for Nanoscale Science, Ithaca, New York 14853, USA}

\begin{abstract}
We report measurements of the local temperature-dependent penetration depth, $\lambda(T)$, in the Kagome superconductors AV$_3$Sb$_5$ (A = Cs, K, Rb) using scanning superconducting quantum interference device (SQUID) microscopy. Our results suggest that the superconducting order in  all three compounds is fully gapped, in contrast to reports of nodal superconductivity in KV$_3$Sb$_5$ and RbV$_3$Sb$_5$. Analysis of the temperature-dependent superfluid density, $\rho_s(T)$, shows deviations from the behavior expected for a single isotropic gap, but the data are well described by models incorporating either a single anisotropic gap or two isotropic gaps. Notably, the temperature dependences of $\lambda(T)$ and $\rho_s(T)$ in KV$_3$Sb$_5$ and RbV$_3$Sb$_5$ are qualitatively more similar to each other than to CsV$_3$Sb$_5$, consistent with the superconducting phase reflecting features of the normal-state band structure. Our findings provide a direct comparison of the superconducting properties across the AV$_3$Sb$_5$ family.

\end{abstract}

\maketitle
The Kagome lattice, formed by corner-sharing triangles, leads to geometric frustration, making insulating Kagome materials candidates for exotic magnetic states such as quantum spin liquids \cite{Balents2010,Broholm2020}. In metallic Kagome materials, this lattice motif gives rise to remarkable electronic phenomena such as Dirac points, van Hove singularities, and flat bands, which can result in topologically nontrivial and strongly correlated states of matter. The recently discovered vanadium-based Kagome metals AV$_3$Sb$_5$ (A = Cs, K, Rb) feature alternating V$_3$Sb$_5$ Kagome sheets separated by layers of alkali metal ions, which donate electrons to the Kagome layers\cite{Ortiz2020,Yin2021a,ortiz2021b}. These materials exhibit similar band structures near the Fermi level with multiple Fermi surfaces, dominated by vanadium d-orbitals, and host rich phase diagrams with a charge-density wave (CDW) phase featuring giant anomalous Hall effects, nematicity, and time-reversal symmetry breaking, and superconductivity emerging at low temperatures \cite{Yang2020,Yu2021,Wang2023AH,Nie2022,Guo2022,Khasanov2022,Guguchia2023,Mielke2022,JiangReview2022,yinReview2022,ZhouReview2024}. A key question is clarifying the symmetry of the superconducting order parameter and its relationship with the CDW phase.

Despite their structural and electronic similarities, the three AV$_3$Sb$_5$ compounds display subtle yet significant differences. The CDW distortions CsV$_3$Sb$_5$ differ from those in KV$_3$Sb$_5$ and RbV$_3$Sb$_5$, likely resulting from subtle differences in the ordering of Van Hove singularities near the Fermi level, which are caused by slightly different lattice constants in CsV$_3$Sb$_5$ compared to KV$_3$Sb$_5$ and RbV$_3$Sb$_5$ \cite{Ortiz2021,Kang2022}. In addition, the phase diagram including the CDW phase and superconductivity as a function of pressure and doping in CsV$_3$Sb$_5$ differs from that of the other two compounds \cite{WilsonReview2024,ZhouReview2024}.

These distinctions raise the question of how similar the superconducting phases are across the AV$_3$Sb$_5$ compounds and whether they exhibit different superconducting gap structures. The temperature dependence of the superconducting penetration depth, $\lambda(T)$,  is sensitive to the magnitude of the superconducting gap, $\Delta(\textbf{k})$, and can reveal the presence of nodes \cite{Prozorov2006}. By determining whether nodes are present in the gap, the possible pairing symmetry can be constrained. 

The temperature dependence of the penetration depth in CsV$_3$Sb$_5$ has been probed using a variety of techniques, including tunnel-diode oscillator, $\mu$SR, critical field measurements, and critical current measurements \cite{Duan2021,Ni2021,Roppongi2022,Li2022,Shan2022,Gupta2022,Gupta2022Apr,Zhang2023,Grant2024}. These studies consistently suggest that the superconducting order parameter in CsV$_3$Sb$_5$ is fully gapped across the Fermi surface. 
The number of studies of $\lambda(T)$ in KV$_3$Sb$_5$ and RbV$_3$Sb$_5$ is more limited, and existing studies yield partially conflicting results. Measurements in KV$_3$Sb$_5$ and RbV$_3$Sb$_5$ using $\mu$SR suggested the presence of nodes in KV$_3$Sb$_5$ and RbV$_3$Sb$_5$ at ambient pressure \cite{Guguchia2023}. However, point-contact spectroscopy on KV$_3$Sb$_5$ and CsV$_3$Sb$_5$ indicated a fully gapped superconducting state in both \cite{Yin2021b}.
To resolve these ambiguities and gain deeper insights into the superconducting phase in the AV$_3$Sb$_5$ family, we report a direct, side-by-side comparison of the penetration depth measured by a consistent experimental approach across the compounds. Specifically, we use scanning superconducting quantum interference device (SQUID) susceptometry to systematically measure the temperature-dependent penetration depth for each of the Kagome superconductors AV$_3$Sb$_5$ (A = Cs, K, Rb).

\begin{figure*}[htbp]
  \centering
    \includegraphics[width=1\textwidth]{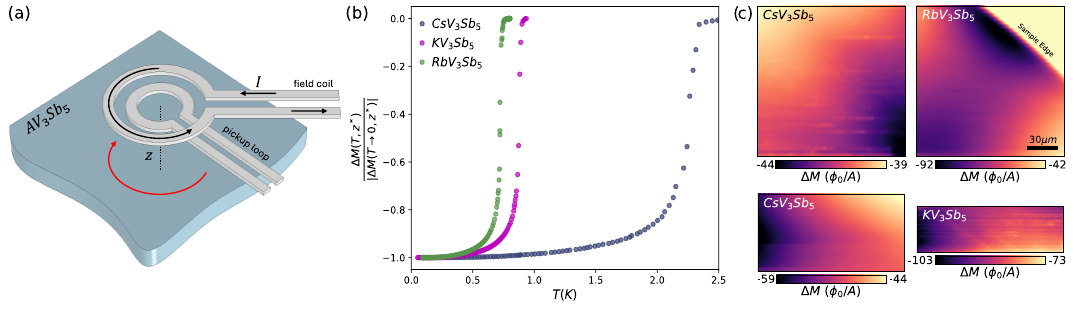}
    \caption{(a) Schematic of the SQUID pickup loop with concentric field coil above the sample. An AC current through the field coil generates  a local magnetic field, which is screened by currents in the superconductor. This alters the magnetic flux in the pick-up loop, detected as a change in mutual inductance between the two coils. (b) Temperature dependence of the change in mutual inductance, $\Delta M(T)$, at a fixed SQUID-sample distance $z=z^*$. $\Delta M(T)$ is normalized by its lowest temperature value for each curve. 
    $\Delta M(T)$ remains zero for $T>T_c$ and decreases sharply at $T_c=2.33$\,K, $0.89$\,K, $0.74$\,K for CsV$_3$Sb$_5$, KV$_3$Sb$_5$, and RbV$_3$Sb$_5$ respectively. (c) Spatial maps of $\Delta M$ on different samples as labeled.}
    \label{fig1}
\end{figure*}

Our SQUID susceptometer consists of two concentric coils: an inner pickup loop which is connected to the SQUID circuit to detect magnetic flux, and an outer field coil, through which an AC current is applied to generate a local magnetic field at the sample \cite{Huber2008}.
The two coils are few $\mu$m in dimensions and we achieve a spatial resolution of $\sim$2-3$\mu$m (See Supp. Sec. \ref{sec:SQUID} for details).
When the SQUID is far from the sample, the mutual inductance between these two coils, $M_0$, is measured. As the SQUID approaches a superconducting sample, screening currents in the superconductor reduce the local magnetic field produced by the field coil (Fig.\ref{fig1}(a)) and, consequently, results in a decrease in the mutual inductance, $\Delta M$. $\Delta M$ at a distance $z$ between the SQUID and the sample can be modeled by \cite{kirtley2012hsweep}:
\begin{equation}
    \centering
    {\frac{\Delta M(z,T)}{M_0} = -\frac{1}{(1+\frac{4}{a^2}(z+\lambda(T))^2)^{3/2}}}
    \label{eqn:approachcurve}
\end{equation}
where $a$ is the effective field coil radius, and $\lambda$ is assumed much smaller than the thickness of the sample. We use a Helmholtz coil to compensate for small out-of-plane background magnetic fields (see Supp. Sec. \ref{sec:field_compensation} for details).

Fig.\ref{fig1}(b) shows $\Delta M$ as a function of temperature, normalized by its value at the lowest temperature to allow comparison across different measurements. Measurements were taken with the SQUID in light mechanical contact to maintain a constant height $z=z^*$, with additional out-of-contact data provided in the supplementary material \cite{supp}. The exact value of $z^*$ is not precisely known, and likely differs between measurements due to variations in the SQUID alignment. As temperature decreases, we observe the onset of diamagnetism, indicated by $\Delta M<0$, at the critical temperature $T_c$, followed by an increase in the magnitude of $\Delta M$, reflecting a growing superfluid density. We determine $T_c$ as the temperature at which $\Delta M$ first becomes negative followed by a sharp increase in $|\Delta M|$ to be $2.33$\,K, $0.89$\,K, and $0.74$\,K for CsV$_3$Sb$_5$, KV$_3$Sb$_5$, and RbV$_3$Sb$_5$, respectively. At this temperature, screening currents over the length scale of the field coil are present which typically coincides with the temperature where resistance vanishes in a transport measurement.

Fig.\ref{fig1}(c) reveals spatial variation in $\Delta M$ reflecting variations in the sample's diamagnetic response. Some of this variation is due to changes in the scan height above the non-planar sample surface, but part reflects inhomogeneity of the superfluid density, which we verify by observing that vortices preferentially form in regions with smaller $|\Delta M|$ when we cool the sample through $T_c$ in a small magnetic field (Supp. Sec. \ref{sec:inhomo_rho}). The origin of the local variations in the superfluid density is unclear. Possible causes include variations in strain, stoichiometry, or defect density. Despite the non-uniformity, we observe a sharp onset of diamagnetism at $T_c$ (Fig.\ref{fig1}(b)) because we average over only a small volume of the sample ($\sim$ 10\,$\upmu$m$^3$). In contrast, in a bulk measurement that averages over the entire sample, we expect that the observed non-uniformity results in a broadened transition consistent with what has been observed \cite{Duan2021,Roppongi2022,Grant2024}.

\begin{figure}[htbp]
    \includegraphics[width=.48\textwidth]{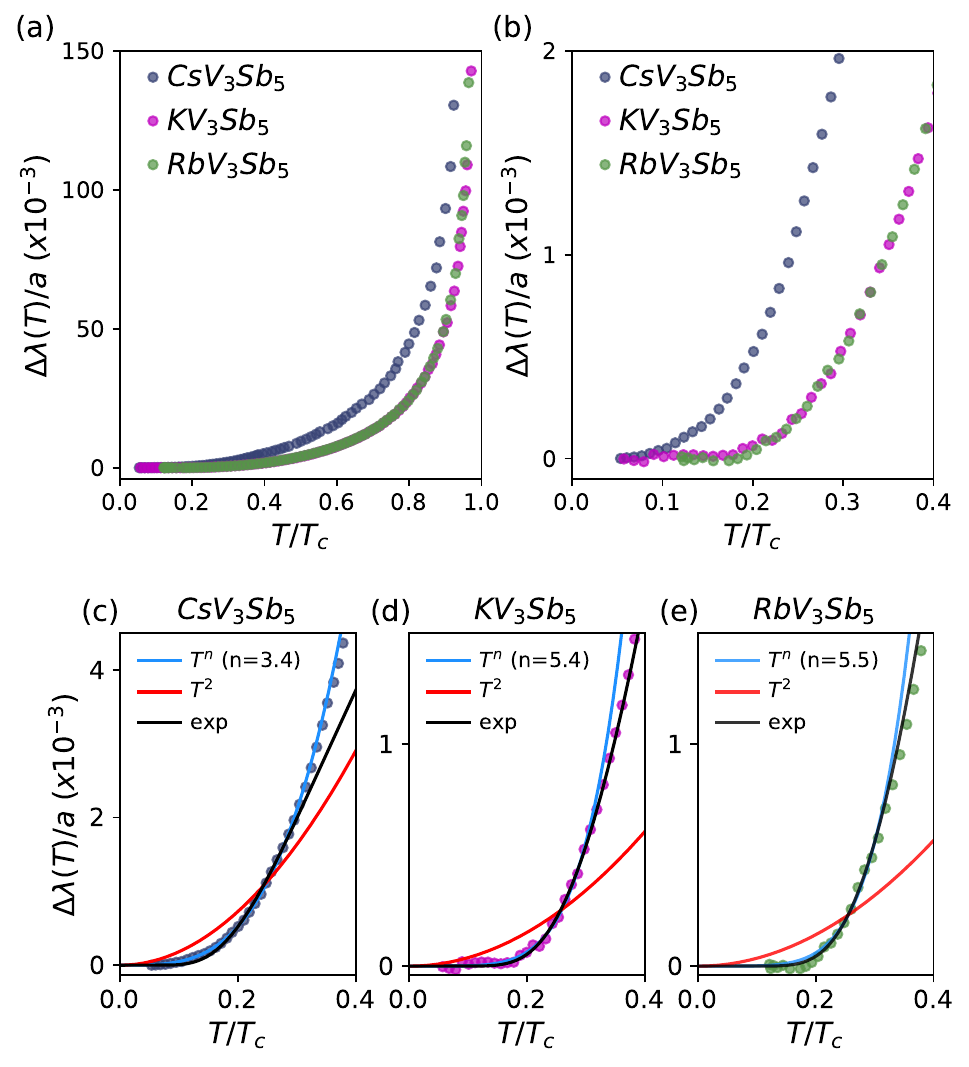}
\caption{Change in penetration depth $\Delta \lambda(T)$ rescaled with the field coil radius $a$ over the (a) full and (b) low temperature range. (c)-(e) Fits of $\Delta \lambda(T)/a$ for each compound as indicated up to $0.3T_c$. Fits to $T^2$ fail to capture the data, and power law fits to $T^n$ yield $n > 2$. Exponential fits (Eq.\ref{eqn:exp}) agree with the data with $\Delta_0$ given by $0.21$\,meV, $0.12$\,meV, $0.10$\,meV for CsV$_3$Sb$_5$, KV$_3$Sb$_5$, and RbV$_3$Sb$_5$ respectively.}
\label{fig:fig2}
\end{figure}

\begin{figure*}[htbp]
  \centering
    \includegraphics[width=\textwidth]{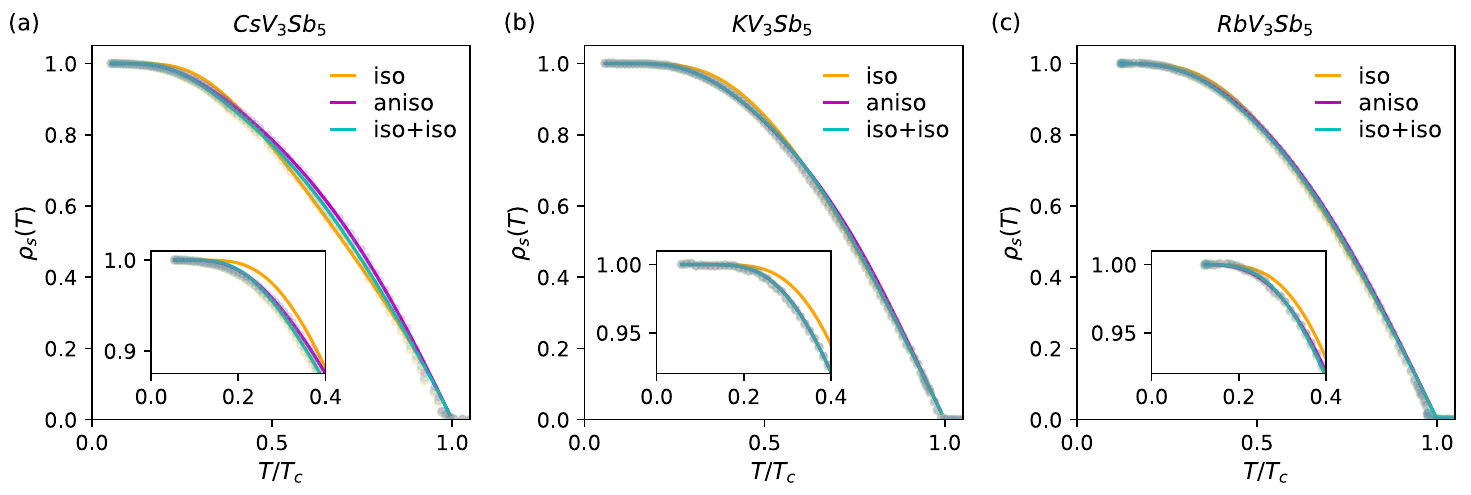}
  \caption{ Reduced superfluid density for (a) CsV$_3$Sb$_5$, (b) KV$_3$Sb$_5$, and (c) RbV$_3$Sb$_5$. In each panel, the three sets of data points are calculated from $\Delta M(T)$ using the different values of $\lambda_0/a$ from fitting to three models of the superconducting gap: single isotropic (iso), single anisotropic (aniso), and two isotropic (iso+iso). Similar values of $\lambda_0/a$ result in nearly overlapping data points. Lines show $\rho_s(T)$ corresponding to the best fit for the same models. Insets zoom in on the low temperature regime where the differences between the fits most significant.}
  \label{fig3}
\end{figure*}

At a fixed SQUID-sample distance $z=z^*$, we estimate $\Delta \lambda(T)=\lambda(T)-\lambda_0$ from the measured $\Delta M$, where $\lambda_0$ is the zero temperature penetration depth. From Eq.\ref{eqn:approachcurve}, we see that 
\begin{equation}
y(z^*,T) = \frac{1}{2}\sqrt{\left(-\frac{M_0}{\Delta M(z^*,T)}\right)^{2/3}-1 } = \frac{1}{a}(z^*+\lambda(T)),
\end{equation}
where we obtain the dimensionless quantity $y(z^*,T)$ directly from our data. With this we can write 
\begin{equation}
        \Delta \lambda(T) = a \cdot(y(z^*,T) - y(z^*,T\rightarrow0))
    \label{eqn:deltalambda}
\end{equation}
The extracted value for $\Delta \lambda(T)$ is independent of the exact height $z^*$. The main source of uncertainty is the effective field coil radius, $a$, due to the field coil's finite trace width and deviations from an ideal circular shape. 

In Fig.\ref{fig:fig2}, we present the extracted $\Delta \lambda(T)/a$ for the three AV$_3$Sb$_5$ compounds over the full temperature range (Fig.\ref{fig:fig2}(a)) and at low temperatures (Fig.\ref{fig:fig2}(b)). At low temperatures ($T \lesssim 0.3 T_c$), where the superconducting gap is generally considered to be temperature-independent, $\Delta\lambda(T)$ can reveal the nodal structure of the gap. If nodes are present, $\Delta\lambda(T)$ should follow a power law, with an exponent smaller than 2 that depends on the dimensionality of the node and disorder in the superconductor \cite{Prozorov2006}. In contrast, in the absence of nodes, $\Delta\lambda(T)$ should exhibit an exponential temperature dependence:
\begin{equation}
        \Delta \lambda(T) \propto T^{-1/2}e^{-\frac{\Delta_0}{k_BT}} 
        \label{eqn:exp}
\end{equation}

From Fig.\ref{fig:fig2}(b), we observe that $\Delta \lambda(T)$ saturates at low temperature  for all three compounds. In Fig.\ref{fig:fig2}(c)-(e), we fit $\Delta \lambda(T)/a$ for each compound using Eq.\ref{eqn:exp} for $T < 0.3T_c$, showing good agreement with the data. The fitted gap energies $\Delta_0$ are $1.05, 1.56, 1.57$\,$k_B$$T_c$ for CsV$_3$Sb$_5$, KV$_3$Sb$_5$, and RbV$_3$Sb$_5$, respectively. These values are lower than the BCS weak coupling value of 1.76\,$k_B$$_c$, with CsV$_3$Sb$_5$ showing a significant deviation, and KV$_3$Sb$_5$ and RbV$_3$Sb$_5$ are nearly identical. Fits with a $T^2$ dependence, which would correspond to the presence of line nodes \cite{Prozorov2006} fail to capture the data. In addition, we fit a $T^n$ dependence with $n$ as a fit parameter. The required values of $n$ to obtain reasonable fits are well above $n=2$, and ncrease as we gradually reduce the maximum temperature included in the fit (see Supp. Sec. \ref{sec:Tmax_fitting}). Taken together this analysis suggests that the superconducting gap is nodeless in all three compounds.

The value of the gap from fitting to Eq.\ref{eqn:exp} is sensitive to the selected temperature range. If we choose a temperature cut-off comparable to Roppongi et al. \cite{Roppongi2022} and Duan et al. \cite{Duan2021} ($T/T_c < 0.2$), we find $\Delta_0$ = $0.12$\,meV = $0.60$\,$k_B$$T_c$ for  CsV$_3$Sb$_5$ which is consistent with their values. However, if we choose a larger temperature range ($T/T_c < 0.3$), we find that the gap extracted from our data is higher than the gaps reported in Roppongi et al. and Duan et al. (see Sepp. Sec. \ref{sec:Tmax_fitting} an \ref{sec:lit_compare} for detailed discussion and fitting of our data and data from Ref. \cite{Duan2021,Roppongi2022}). 

For CsV$_3$Sb$_5$, our findings align with other penetration depth measurements, supporting that CsV$_3$Sb$_5$ is fully gapped \cite{Duan2021,Roppongi2022,Yin2021b,Gupta2022,Zhang2023,Grant2024}. For KV$_3$Sb$_5$, our results are consistent with point-contact spectroscopy measurements suggesting nodeless superconductivity \cite{Yin2021b}, though for both KV$_3$Sb$_5$ and RbV$_3$Sb$_5$, our findings contrast with prior penetration depth studies suggesting nodal superconductivity at ambient pressure \cite{Guguchia2023}. 

Our measurements were performed consistently using SQUID sensors with identical dimensions, allowing a direct comparison of $\Delta \lambda(T)/a$ across samples, despite uncertainties in $a$. We find that the shape and magnitude of $\Delta \lambda(T)$ for KV$_3$Sb$_5$ and RbV$_3$Sb$_5$ are nearly identical, while CsV$_3$Sb$_5$ behaves differently. A comparison of the AV$_3$Sb$_5$ series in the normal state from quantum oscillations measurements suggests much closer similarities between the band structures of KV$_3$Sb$_5$ and RbV$_3$Sb$_5$ compared to CsV$_3$Sb$_5$ \cite{Wang2023}. In addition, the CDW distortions in KV$_3$Sb$_5$ and RbV$_3$Sb$_5$ show a 2x2x2 reconstruction while CsV$_3$Sb$_5$ shows a mixed-phase 2x2x4 reconstruction \cite{Ortiz2021,Kang2022, PhysRevMaterials.8.093601}. The close similarities between the superconducting phases in KV$_3$Sb$_5$ and RbV$_3$Sb$_5$ compared to CsV$_3$Sb$_5$ in our measurements are aligned with these studies showing KV$_3$Sb$_5$ and RbV$_3$Sb$_5$ are more similar in the normal state than CsV$_3$Sb$_5$. Together this indicates the superconducting state inherits features of the Fermiology of the normal state.

Next, we analyze our data for the three AV$_3$Sb$_5$ compounds across the full temperature range, focusing on the reduced superfluid density, $\rho_s(T) = \lambda_0^2/\lambda^2(T)$. The temperature dependence of $\rho_s$ is typically compared to semiclassical models that incorporate the temperature dependence of the gap over the Fermi surface, $\tilde{\Delta}(T,\mathbf{k})$ \cite{Prozorov2006}. Previous studies incorporating this analysis have suggested anisotropic and/or multiple superconducting gaps contributing to the superfluid density in CsV$_3$Sb$_5$ \cite{Duan2021,Roppongi2022,Ni2021,Gupta2022,Gupta2022Apr,Xu2021,Yin2021b,Grant2024,Shan2022,Li2022,Roppongi2022}. Here, we perform a similar analysis of $\rho_s(T)$ for our data across the AV$_3$Sb$_5$ series. Specifically, we attempt to fit our data to models of $\rho_s(T)$ for a single isotropic gap, a single anisotropic gap, and two isotropic gaps. Details of the different models of $\rho_s(T)$ can be found in Supp. Sec. \ref{sec:rho_models}. The $\rho_s(T)$ corresponding to these different models of the gap show only subtle differences in their curvature and shape. At the same time, a key challenge is the need to assume a value for $\lambda_0$ to calculate $\rho_s(T)$ from the data. While changes in the penetration depth can be reliably measured using various techniques, determining the absolute value of $\lambda_0$ is more difficult.  
These factors make it inherently challenging to distinguish between a single isotropic gap, an anisotropic gap, and/or the presence of multiple gaps.

In the following, we fit our measured $\Delta M$ data directly using $\lambda_0/a$ as a fit parameter rather than assuming a value. We then compare the quality of the fits across different models of the gap and the resulting values for $\lambda_0/a$. Our fitting procedure uses the following expression based on  Eq.\ref{eqn:deltalambda}: 
\begin{equation}
    \frac{\Delta M(T)}{M_0} = -\left(1+4\left[y(0) + \frac{\lambda_0}{a}(\rho_s(T)^{-1/2}-1)\right]^2\right)^{-3/2}
    \label{eqn:M_rho}
\end{equation}

In Fig.\ref{fig3}, we show the best-fit $\rho_s(T)$, alongside the $\rho_s(T)$ extracted from the measured $\Delta M(T)$ using the fitted values of $\lambda_0/a$ and inverting Eq.\ref{eqn:M_rho}. Fits to $\Delta M(T)$ and the corresponding fit parameters for each compound and gap model are shown in the supplementary material (Supp. Sec. \ref{sec:deltaM_fit} and \ref{sec:fit_params}). The values of $\lambda_0/a$ are primarily constrained by $\Delta M(T)$ and differ by less than approximately $10\%$ across the different gap models for a given compound. Bounding the value of $\lambda_0$ based on the maximum uncertainty in $a$, we find $\lambda_0$ = $195$-$390$\,nm, $127$-$255$\,nm, $123$-$247$\,nm for CsV$_3$Sb$_5$, KV$_3$Sb$_5$, and RbV$_3$Sb$_5$ respectively. For CsV$_3$Sb$_5$ this is consistent with the value of $\lambda_0$ estimated in \cite{Duan2021}.
 
Across the full temperature range, the gap models capture the data effectively (Fig.\ref{fig3}). The insets in Fig.\ref{fig3} highlight low temperatures where differences between the fits are the most pronounced. For all three compounds, a single isotropic gap shows discrepancies from the data, consistent with previous reports on CsV$_3$Sb$_5$ \cite{Duan2021,Roppongi2022}. The gaps and fit parameters obtained from the fitting are summarized in Supp. Sec. \ref{sec:fit_params}. The minimum of the gap in the anisotropic model and the smaller gap in the multigap model determine the behavior of the superfluid density at  low temperature. For all compounds, the smaller gap in the multigap model aligns with the gap extracted from exponential fits to the low-temperature behavior in Fig.\ref{fig:fig2}. In the anisotropic model, the minimum of the gap is smaller, likely to assign enough angular weight at low gap values to capture the low-temperature behavior. The average gap values from the anisotropic and multigap models are consistent. In RbV$_3$Sb$_5$, the larger gap in the multigap model takes on a high value, but this is offset by a low weighting factor, suggesting strong correlation between these fit parameters and limited constraint on the larger gap value.

Comparing the fit results across the three compounds, we find that RbV$_3$Sb$_5$ and KV$_3$Sb$_5$ show similar trends and values, consistent with the similarity in their curves, while CsV$_3$Sb$_5$ behaves distinctly. For both the anisotropic and multigap models, more weight is effectively assigned to the larger gap for CsV$_3$Sb$_5$, reflected in a higher weighting factor in the multigap fit and greater anisotropy in the anisotropic model. In Supp. Sec. \ref{sec:dlam_rescale}, we perform a more direct comparison of the data without fitting by rescaling the measured $\Delta\lambda(T)$ to eliminate the influence of the gap magnitude and the value of $\lambda_0$ on its curvature. This analysis suggests that differences in $\Delta\lambda(T)$ measured in CsV$_3$Sb$_5$ compared to KV$_3$Sb$_5$ and RbV$_3$Sb$_5$ cannot be solely explained by differences in $\lambda_0$ or the ratio between the gap magnitude and $T_c$. This indicates a difference in the structure of the gap for CsV$_3$Sb$_5$ compared to KV$_3$Sb$_5$ and RbV$_3$Sb$_5$, such as differences in the gap anisotropy or a different weighting between multiple gaps, as is observed in the results from fitting to the different gap models.

For all compounds, both the single anisotropic and the two isotropic gap models are nearly indistinguishable. Previous studies have suggested gap anisotropy, multiple gaps and a combination of both \cite{Duan2021,Roppongi2022,Ni2021,Gupta2022,Guguchia2023,Gupta2022Apr,Xu2021,Yin2021b,Grant2024,Shan2022,Li2022}. Here, we are unable to distinguish between these scenarios from the temperature dependence of the superfluid density. We believe that this conclusion is not unexpected. The penetration depth reflects the total superfluid density, which quantifies the number of condensed electrons while being agnostic to their origin in momentum space. The difference between a multigap and anisotropic gap model lies in their effective distribution of gap values: the multigap model uses two discrete gap values with a relative weighting factor, while the anisotropic model describes a continuous distribution of gap values shaped by the sinusoidal anisotropy. The measured data does not sufficiently constrain the details of the gap distribution to clearly distinguish between the two possibilities and even more complicated ones.

In summary, we have measured the local temperature-dependent penetration depth of AV$_3$Sb$_5$ (A = Cs, K, Rb) using scanning SQUID microscopy. While some reports suggest nodal superconductivity in KV$_3$Sb$_5$ and RbV$_3$Sb$_5$ and the absence of nodes in CsV$_3$Sb$_5$, our findings indicate that all three AV$_3$Sb$_5$ compounds are fully gapped. Analysis of the temperature-dependent superfluid density,$_s (T)$from the behavior expected for a single isotropic gap for all three  compounds. However, our data are well described by models incorporating either a single anisotropic gap or two isotropic gaps, making it challenging to determine which scenario is more likely. The temperature dependences of the penetration depth and superfluid density in KV$_3$Sb$_5$ and RbV$_3$Sb$_5$ are much more similar to each other than to CsV$_3$Sb$_5$. This is consistent with the similarities between the AV$_3$Sb$_5$ compounds in the normal state, and suggests the superconducting phase inherits features of the normal-state band structure. Our measurements provide a direct comparison of the superconducting phases across the AV$_3$Sb$_5$ series, and highlight the need for probes of the superconducting state beyond the magnetic penetration depth to distinguish between different scenarios for the gap structure.

\section*{Acknowledgments}
This work was primarily supported by the Air Force Research Laboratory, Project Grant FA9550-21-1-0429. SDW and ACS gratefully acknowledge support via the UC Santa Barbara NSF Quantum Foundry funded via the Q-AMASE-i program under award DMR-1906325.
\noindent

\bibliography{bib}

\end{document}

\title{Supplemental Material for: Direct Comparison of Magnetic Penetration Depth in Kagome Superconductors AV$_3$Sb$_5$ (A = Cs, K, Rb)}

\date{\today}
\author{Austin R. Kaczmarek}
\affiliation{Laboratory of Atomic and Solid-State Physics, Cornell University, Ithaca, NY, USA}

\author{Andrea Capa Salinas}
\affiliation{Materials Department and California Nanosystems Institute, University of California Santa Barbara, Santa Barbara, CA , USA}

\author{Stephen D. Wilson}
\affiliation{Materials Department and California Nanosystems Institute, University of California Santa Barbara, Santa Barbara, CA , USA}

\author{Katja C. Nowack}
\affiliation{Laboratory of Atomic and Solid-State Physics, Cornell University, Ithaca, NY, USA}

\maketitle

\tableofcontents
\newcounter{somecounter}

\newpage
\stepcounter{somecounter}
\section{Supplementary Note \thesomecounter: Synthesis} \label{sec:synthesis}
Single crystals of AV$_3$Sb$_5$ (A = Cs, K, Rb) were synthesized with the self-flux method. Elemental alkali metal K ingot (Alfa, 99.8$\%$), Rb ingot (Alfa, 99.75$\%$), Cs liquid (Alfa 99.98$\%$), V powder (Sigma 99.9$\%$), and Sb shot (Alfa 99.999$\%$) were weighed out in the 20:15:120 ratio, sealed inside a tungsten carbide vial in an inert environment and milled for 60 minutes in a SPEX 8000D mill. The vanadium powder used was cleaned from oxide impurities with a 10/90 hydrochloric acid and ethanol mixture prior to weighing. The resulting milled precursor powders were loaded into alumina crucibles and sealed inside steel tubes. Further, they were heated up to 1000 $^{\circ}$C, held at that temperature for 12 hours, then cooled down to 900 $^{\circ}$C at 5 $^{\circ}$C/h followed by a cooldown to 500 $^{\circ}$C at 1 $^{\circ}$C/h. Once at room temperature, the resulting crystals were extracted manually in air. Prior to scanning SQUID measurements, the top few layers of the samples were peeled away, revealing clean, mirror-like areas on the sample surface. This was accomplished by wrapping carbon tape over the tip of a toothpick and gently rolling the toothpick on the sample surface, peeling away a few layers at a time. 

\begin{figure}[htbp]
    \centering
    \includegraphics[width =1\columnwidth]{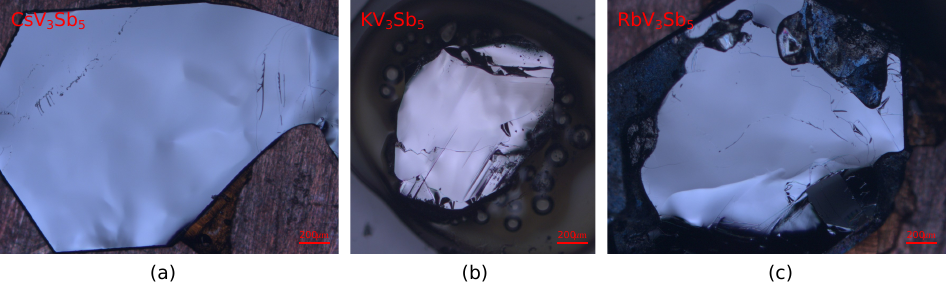}
    \caption{As-grown AV$_3$Sb$_5$ samples. The top few layers have been peeled away with carbon tape, revealing clean areas spanning hundreds of $\upmu$m. In our scanning SQUID measurements, we avoid cracks and defects, and find large, clean, and flat areas to probe.}
    \label{fig:samples_imgs}
\end{figure}
\newpage
\stepcounter{somecounter}
\section{Supplementary Note \thesomecounter: Scanning SQUID} \label{sec:SQUID}

Our SQUID susceptometers include a pickup loop coupled to a SQUID circuit for detecting magnetic flux and a field coil for applying a local magnetic field to the sample. The pickup loop has inner and outer radii of 0.75\,$\upmu$m and 1.6\,$\upmu$m, respectively, while the field coil has inner and outer radii of 3\,$\upmu$m and 6\,$\upmu$m, providing a spatial resolution of $\sim$2-3\,$\upmu$m. The physical limits of the inner and outer radii of the field coil (3\,$\upmu$m and 6\,$\upmu$m respectively) are used to place the bounds on the values of $\lambda_0$ as presented in the main text.

The design of the susceptometers is gradiometric: a second field coil and counter-wound pickup loop are incorporated to minimize the net flux coupled into the SQUID circuit from both a uniform background field and the field applied by the field coil \cite{Huber2008}. When the susceptometer approaches the sample, one pair of the pickup loop and field coil (front) comes close to the surface, while the other pair (rear) stays far from the sample surface. Current through the field coil circuit couples flux with opposite sign into the two counter-wound pickup loops. As a result, far from the sample surface, the measured SQUID signal $S$ reflects the mismatch between the mutual inductances of the two pairs, i.e.  $S(z\gg a)=(M_{front}-M_{rear})I_{FC}$ with $a$ the effective field coil radius and $I_{FC}$ the current in the field coil. As the front pair approaches the sample surface at distance $z$ the sample modifies $M_{front}$ by $\Delta M(z)$, and our total measured signal is $S(z) = (M_{front} + \Delta M(z) -M_{rear})I_{FC}$. In the main tet, we report $\Delta M(z)$, which we obtain by subtracting $S(z\gg a)$ from the signal and dividing by $I_{FC}$. The unmodified mutual inductance from the main text $M_0$ is equal to $M_{front}$, which we  measure by flowing current only through the front field coil and detecting the resulting SQUID signal far from the sample. $M_0$ is approximately 398\,$\upphi_0$/A for the SQUIDs we used in this study.

\newpage
\stepcounter{somecounter}
\section{Supplementary Note \thesomecounter: Modeling of $\rho_s(T)$ for Different Gaps} \label{sec:rho_models}
Here, we describe the models corresponding to different gap structure used to fit the superfluid density in the main text. The field coil applies predominantly an out-of-plane field, i.e. along the c-axis of the sample. The generated screening currents in the superconductor circulate in-plane.  We therefore measure the in-plane penetration depth $\lambda_{ab}$, and model the superfluid density $\rho_{s_{ab}}$ similar to Refs. \cite{Duan2021,Roppongi2022,Ni2021,Gupta2022,Guguchia2023,Gupta2022Apr,Grant2024,Shan2022,Li2022}.

The electronic state of AV$_3$Sb$_5$ is quasi-2D \cite{Wang2023}. As a result, We assume a cylindrical Fermi surface, for which the superfluid density is given by\cite{Prozorov2006}:

\begin{equation}
    \centering
    \rho_{s_{ab}}(T) = 1 - \frac{1}{2\pi k_B T} \int_{0}^{2\pi}\int_{0}^{\infty}cosh^{-2}\left(\frac{\sqrt{\varepsilon^2+\tilde{\Delta}^2(T,\varphi)}}{2k_BT}\right)\,d\varepsilon d\varphi
    \label{eqn:rho(T)}
\end{equation}

where $\tilde{\Delta}$ is the gap and $\varphi$ is the in-plane azimuthal angle.  We introduce in-plane anisotropy of the gap through
$\tilde{\Delta}(T,\varphi) = \Delta(T)f(\varphi)$, where $f(\varphi) = \frac{1+\gamma cos(6\varphi)}{\sqrt{1+\gamma^2/2}}$ reflects the sixfold rotational symmetry of the Kagome lattice, and $0\leq \gamma \leq1$ sets the degree of anisotropy. $f(\varphi)$ is written such that the RMS value of $f(\varphi)$ over the Fermi surface is equal to 1. For an isotropic gap, $\gamma=0$ and $f(\varphi)=1$. For the temperature dependence of the gap, we use the gap interpolation formula:
\begin{equation}
    \centering
    \Delta(T) = \Delta_0 tanh \left(\frac{\pi k_BT_c}{\Delta_0}\sqrt{\frac{T_c}{T}-1}\right).
    \label{eqn:Delta(T)}
\end{equation}

To account for multiple superconducting gaps, we use the phenomenological alpha-model previously applied to the multi-gap superconductor MgB$_2$ \cite{Manzano2002}: 

\begin{equation}
    \centering
     \rho_s(T) = \alpha\rho_{s_1}(\tilde{\Delta}_1(T,\varphi)) + (1-\alpha)\rho_{s_2}(\tilde{\Delta}_2(T,\varphi)),
\end{equation}

where $\alpha$ sets the ratio of contributions from two bands with superfluid densities $\rho_{s_1}$ and $\rho_{s_2}$ and gaps $\Delta_1$ and $\Delta_2$, respectively.

The reduced superfluid density and the penetration depth are related through

\begin{equation}
    \centering
    \rho_s(T) = \frac{n_s(T)}{n_{s}(T=0)} = \left({\frac{\Delta\lambda(T)}{\lambda_0}+1}\right)^{-2}
\label{eqn:rho_lam_relate}
\end{equation}
where $n_s$is the absolute superfluid density and $\lambda_0$ is the zero temperature penetration depth.

\newpage
\stepcounter{somecounter}
\section{Supplementary Note \thesomecounter: Background Field Compensation}
\label{sec:field_compensation}

\begin{figure}[htbp]
    \centering
    \includegraphics[width =.7\columnwidth]{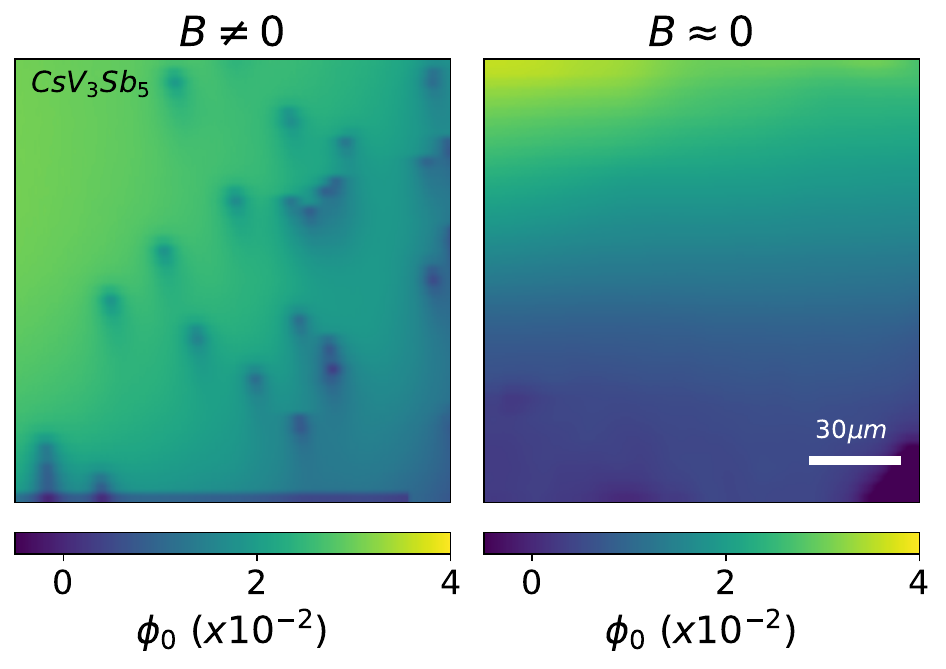}
    \caption{Imaging the sample surface with the SQUID after cooling the sample through $T_c$. With finite background field (left), we see vortices within our field of view. We adjust the background field by changing the current through a Helmholtz-coil surrounding the scanning SQUID microscope. We repeatedly thermal cycle the sample, re-cooling through $T_c$ after each field adjustment until we find $\sim$zero vortices in our field of view (right). This ensures that there are no nearby vortices moving in response to the AC field-coil excitation that could contribute signal in $\Delta M$.}
    \label{fig:field_cool}
\end{figure}

\newpage
\stepcounter{somecounter}
\section{Supplementary Note \thesomecounter: Spatial Inhomogeneity of Superfluid Density} \label{sec:inhomo_rho}
\begin{figure}[htbp]
    \centering
    \includegraphics[width =.9\columnwidth]{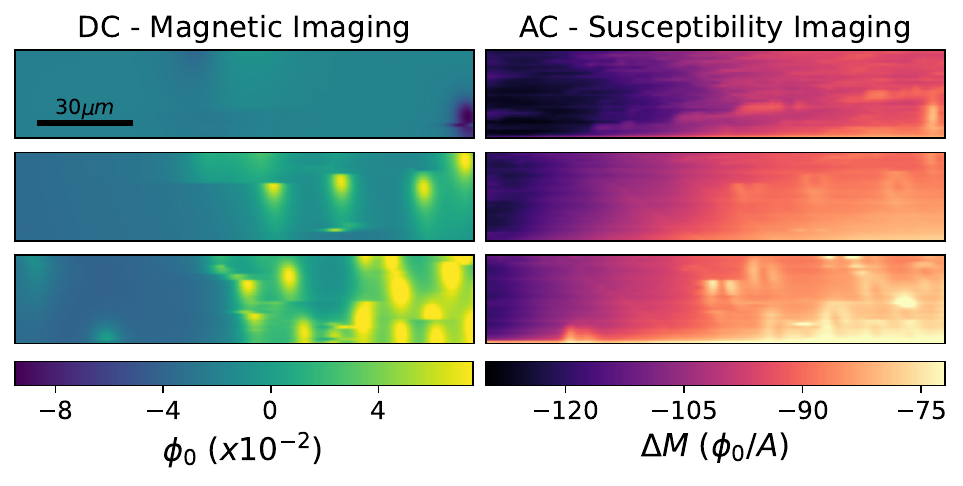}
    \caption{Scanning SQUID imaging of example region in KV$_3$Sb$_5$ showing spatial inhomogeneity in the superfluid density. (left) DC magnetic imaging revealing the DC magnetic flux from vortices. (right) AC susceptibility imaging revealing spatial variation of the diamagnetic response of the superconductor. The sample is cooled through $T_c$ in increasing out-of-plane magnetic field from top to bottom inducing an increasing number of vortices. Vortices preferentially nucleate in the regions which exhibit a weaker diamagnetic response in the susceptibility imaging. This indicates that the regions of weaker diamagnetic response corresponds to reduced superfluid density and that the superfluid density has some inhomogeneity across the sample.}
    \label{fig:inhomo_rho}
\end{figure}

During imaging, we simultaneously record the static (DC) magnetic flux through the pickup loop and the AC susceptibility, $\Delta M$. The DC signal  reveals magnetic features such as Meissner screening and vortices, while the AC imaging probes the diamagnetic response of the superconductor. Spatial variation in the diamagnetic response can be a result of real-space inhomogeneities in the superfluid density or from imaging artifacts such as varying scan height due to a non-planar sample surface. To distinguish between these scenarios, we adjust the magnetic field using a Helmholtz coil surrounding the SQUID microscope and cool the sample through $T_c$ under increasing magnetic field strength. Cooling through $T_c$ in a finite field induces vortices, which preferentially nucleate at regions of reduced superfluid density. In Fig.\ref{fig:inhomo_rho}, we observe that vortices preferentially nucleate in regions with a lower diamagnetic susceptibility (lighter regions in right column). This indicates that the  spatial variation observed in AC susceptibility reflects some amount of real-space inhomogeneity in the superfluid density.

\newpage
\stepcounter{somecounter}
\section{Supplementary Note \thesomecounter: Rescaling of $\Delta\lambda$} \label{sec:dlam_rescale}

\begin{figure}[htbp]
    \centering
    \includegraphics[width =1\columnwidth]{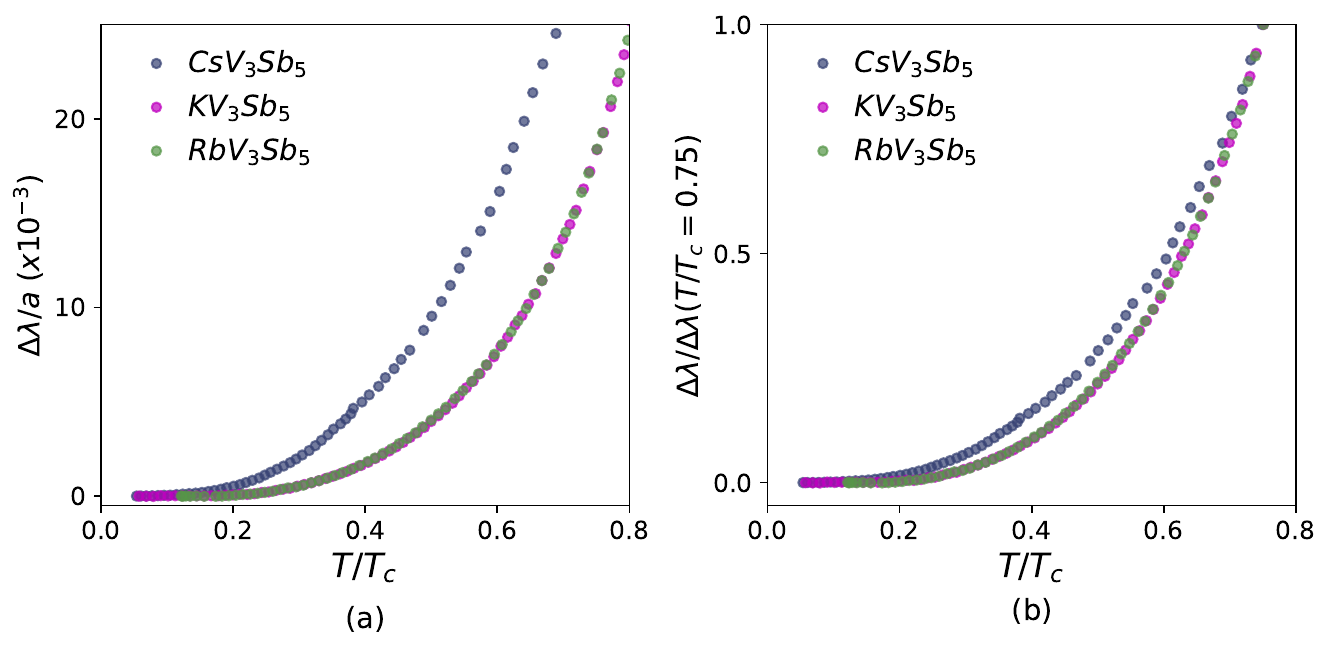}
    \caption{(a) $\Delta\lambda/a$ vs. $T/T_c$ plotted for the three AV$_3$Sb$_5$ compounds (same as Fig.\ref{fig:fig2}(a) and (b) in the main text). (b) $\Delta\lambda$ vs. $T/T_c$ rescaled by its value at 0.75$T/T_c$. CsV$_3$Sb$_5$ shows a different curvature than KV$_3$Sb$_5$ and RbV$_3$Sb$_5$, indicating the difference in $\Delta\lambda$ cannot be explained only by a difference in $\lambda_0$.}

    \label{fig:dlam_rescale}
\end{figure}

From Eq.\ref{eqn:rho_lam_relate}, we can write $\Delta\lambda(T) = \lambda_0\left(\frac{1}{\sqrt{\rho_s(T)}}-1\right)$. $\rho_s(T)$ is the reduced superfluid density, and its temperature dependence results from the structure of the gap along with the gap magnitude and is typically  modeled by Eq.\ref{eqn:rho(T)} or similar expressions. Rescaling $T$ by $T_c$ eliminates the influence of the gap magnitude on $\rho_s(T)$ (assuming the gap scales linearly with $T_c$) and differences in $\rho_s(T/T_c)$ should only manifest if there is a difference in the gap structure.
Typical experiments probe $\Delta\lambda(T)$ rather than $\rho_s(T)$. Similarly, rescaling $T$ by $T_c$ eliminates the influence of the gap magnitude on $\Delta\lambda(T)$. However, $\Delta\lambda(T)$ is additionally influenced by the value of $\lambda_0$.  
Rescaling $\Delta\lambda(T)$ by its value at some temperature $\Delta\lambda(T^*)$ eliminates the influence of $\lambda_0$. Comparing the rescaled curves of $\Delta\lambda(T/T_c)/\Delta\lambda((T/T_c)^*)$ should reveal differences only in the gap structure. In Fig.\ref{fig:dlam_rescale}(a) we re-plot $\Delta\lambda(T)/a$ curves for the three AV$_3$Sb$_5$ compounds (same as Fig.\ref{fig:fig2}(a) and (b) in the main text). In Fig.\ref{fig:dlam_rescale}(b) we rescale $\Delta\lambda(T)$ by its value at 0.75$T/T_c$. We find that KV$_3$Sb$_5$ and RbV$_3$Sb$_5$ are nearly overlapping as before, while CsV$_3$Sb$_5$ still shows a different curvature. This indicates that the difference in $\Delta\lambda(T)$ measured in CsV$_3$Sb$_5$ compared to KV$_3$Sb$_5$ and RbV$_3$Sb$_5$ cannot be explained solely by differences in $\lambda_0$ and the gap magnitude and suggests a difference in the structure of the gap such as e.g. a different gap anisotropy or a different weighting between multiple gaps as is observed in the results from fitting to the different gap models.

\newpage
\stepcounter{somecounter}
\section{Supplementary Note \thesomecounter: Sample-to-Sample Variation} \label{sec:Cs_Samples} 
\begin{figure}[htbp]
    \centering
    \includegraphics[width =1\columnwidth]{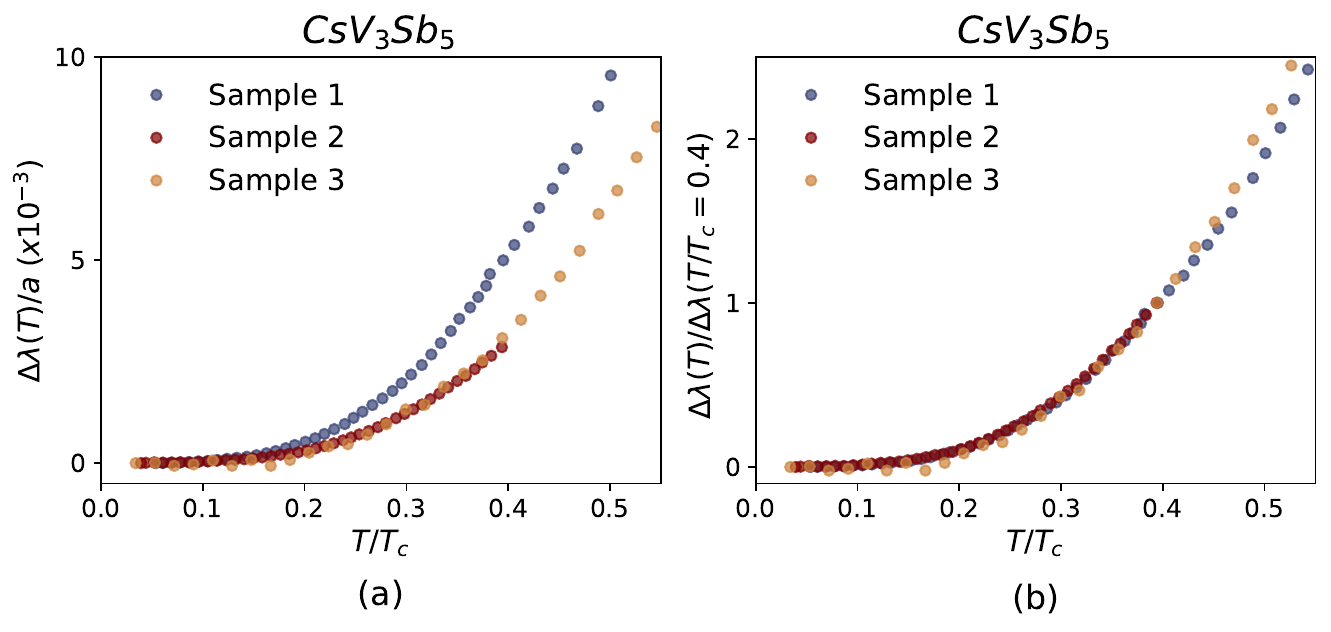}
    \caption{(a) Measurements of $\Delta \lambda$  vs. $T/T_c$ for three different samples of CsV$_3$Sb$_5$. Sample 2 and 3 agree quite well and have similar transition temperatures $T_c=2.62$\,K and $2.64$\,K respectively. Sample 1 has $T_c=2.33$\,K and $\Delta \lambda$ scales a bit different from the other two. Data for samples 1 and 2 are taken with the SQUID in light mechanical contact with the sample. Data for sample 3 is taken with the SQUID out-of-contact from the sample. (b) Rescaling $\Delta \lambda/a$ by its value at $T/T_c = 0.4$ collapses the three curves such that they agree well over this temperature range. As described in Supp. Sec. \ref{sec:dlam_rescale}, rescaling $\Delta \lambda$ and plotting vs $T/T_c$ eliminates the influence of the gap magnitude and $\lambda_0$ on the data and any remaining differences should be indicative of differences in the structure of the gap. The data collapsing in this way indicates the same gap structure in all three CsV$_3$Sb$_5$ samples.
    }
    \label{fig:Cs_Samples}
\end{figure}

\newpage
\stepcounter{somecounter}
\section{Supplementary Note \thesomecounter: Spatial Variation Within the Same sample} \label{sec:Rb_pos} 
\begin{figure}[htbp]
    \centering
    \includegraphics[width =.85\columnwidth]{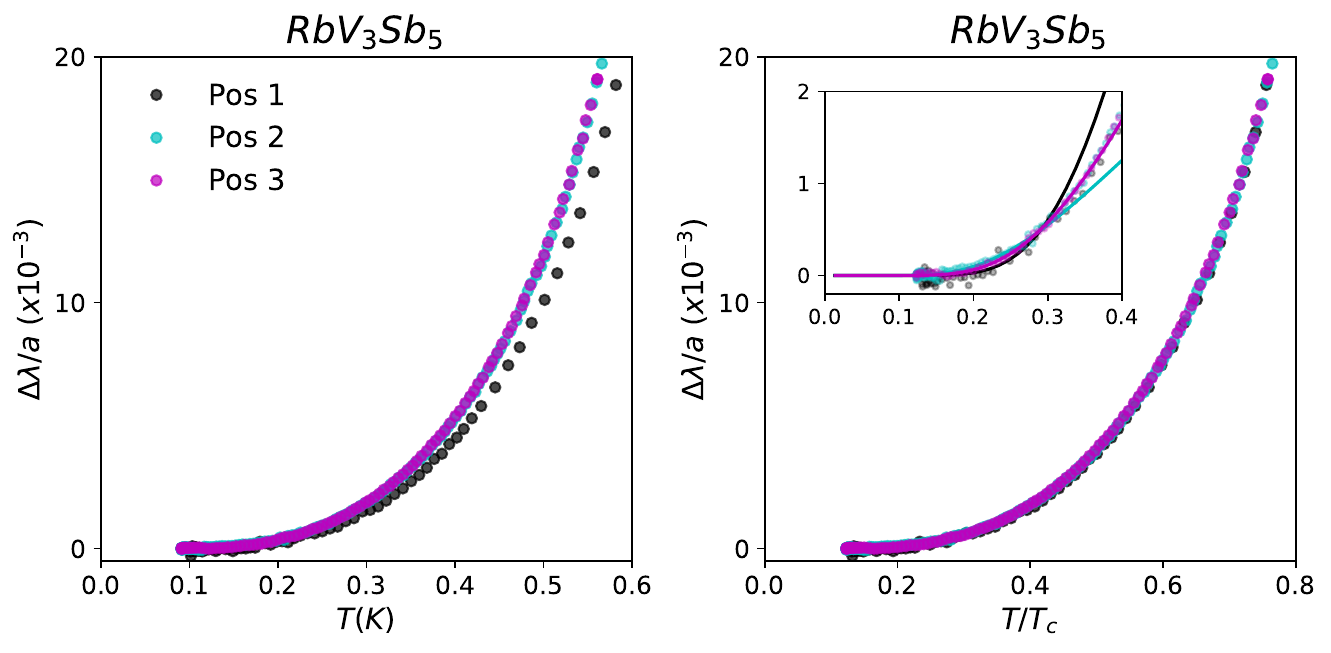}
    \caption{(left) $\Delta \lambda(T)$ curves for three different positions on the same RbV$_3$Sb$_5$ sample. We find slight variations in $T_c$ within the same sample around a few tens of mK, and positions 1 ,2 ,and 3 have $T_c=0.77, 0.74, 0.74$\,K respectively. (right) Plotting $\Delta \lambda(T)$ vs $T/T_c$ collapses the curves onto one another. The inset shows a zoom-in to the low temperature regime along with fits up to $T/T_c=0.3$ to the exponential expression given in Eq.\ref{eqn:exp}.}
    \label{fig:Rb_pos}
\end{figure}

\newpage
\stepcounter{somecounter}
\section{Supplementary Note \thesomecounter: Comparison to Published Data} \label{sec:lit_compare}

\begin{figure}[htbp]
    \centering
    \includegraphics[width =1\columnwidth]{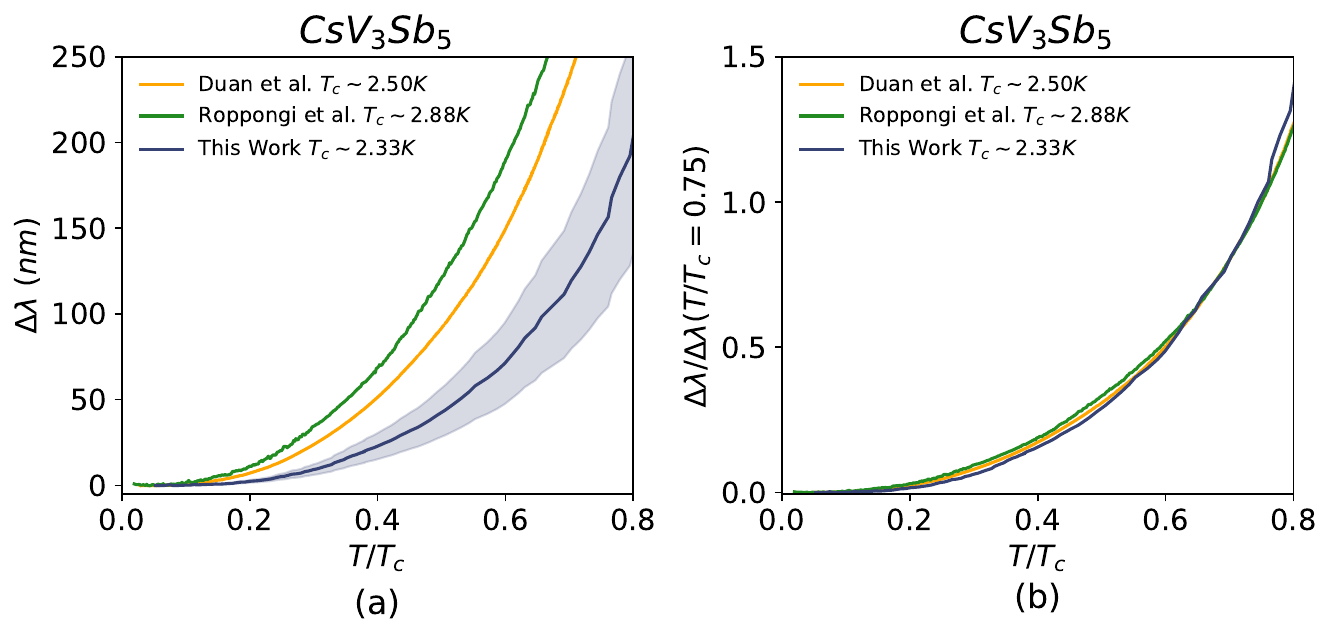}
    \caption{
    (a) Our data plotted alongside the as-published data from Duan et al.\cite{Duan2021} and Roppongi et al.\cite{Roppongi2022} using a tunnel-diode-oscillator (TDO) technique. The shaded region around our data represents the uncertainty in the field coil radius $a$, where the whole $\Delta\lambda(T)$ curve shifts up or down based on the chosen value of $a$. The solid curve corresponds to $a=4.5$\,$\upmu$m. Though the three data do not overlap, they all show a saturation at low temperature indicating CsV$_3$Sb$_5$ is fully gapped. (b) We rescale $\Delta \lambda(T)$ by  its value at $T/T_c = 0.75$. This leads to all three curves mostly collapsing throughout the temperature range. As discussed in Supp. Sec. \ref{sec:dlam_rescale}, this can be explained if the three samples have very similar gaps and therefore superfluid densities $\rho_s(T)$, but differ in the value of $\lambda_0$.}
    \label{fig:lit_compare}
\end{figure}

\newpage
\stepcounter{somecounter}
\section{Supplementary Note \thesomecounter: Sample Thermalization} \label{sec:warming_cooling} 
\begin{figure}[htbp]
    \centering
    \includegraphics[width =1\columnwidth]{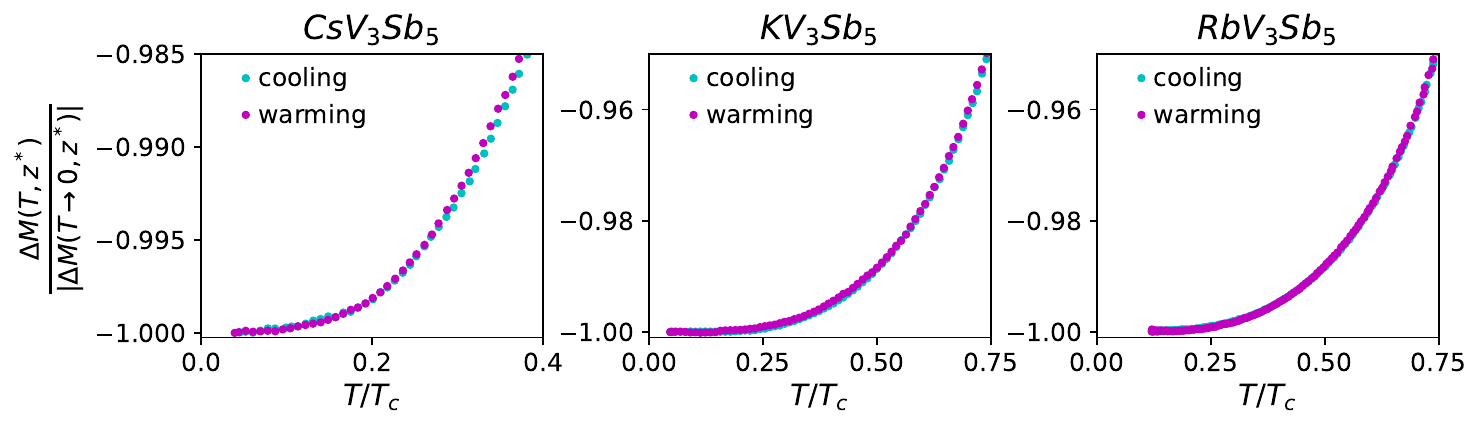}
    \caption{The diamagnetic response as a function of temeprature $\Delta M(T)$ measured while sweeping the sample temperature up and down. The warming and cooling curves are in good agreement, ruling out poor sample thermalization that would result in thermal lag. }
    \label{fig:warming_cooling}
\end{figure}

\newpage
\stepcounter{somecounter}
\section{Supplementary Note \thesomecounter: Fitting $\Delta\lambda$ Over Different Temperature Ranges} \label{sec:Tmax_fitting}

\begin{figure}[htbp]
    \centering
    \includegraphics[width =1\columnwidth]{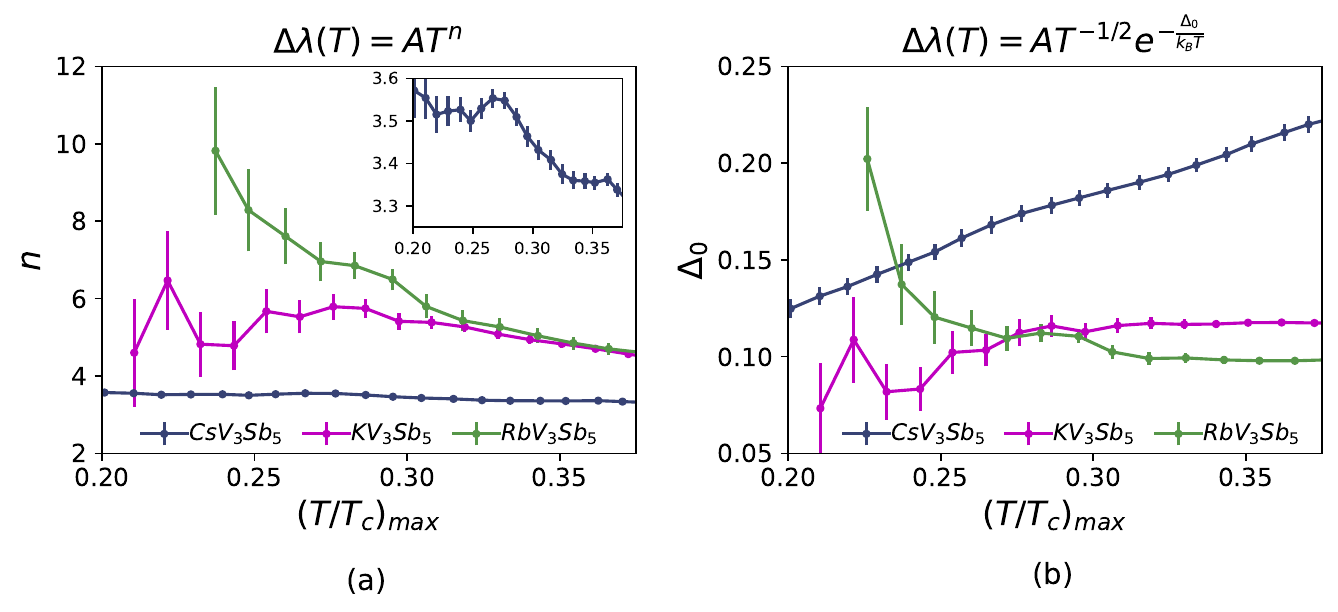}
    \caption{Fig.\ref{fig:fig2} in the main text shows fits to $\Delta\lambda$ in the low temperature regime, fitting $\Delta\lambda$ up to $0.3T_c$ with both a power law dependence ($\Delta\lambda=AT^n$) and an exponential dependence (Eq.\ref{eqn:exp}). Here, we vary the temperature range over which the fitting is done by changing the upper temperature limit of fit $(T/T_c)_{max}$. In Fig.\ref{fig:Tmax_fitting} we show the extracted exponent $n$ from the power law fitting (a) and gap $\Delta_0$ from the exponential fitting (b) as a function of $(T/T_c)_{max}$.}

    \label{fig:Tmax_fitting}
\end{figure}

\newpage
\stepcounter{somecounter}
\section{Supplementary Note \thesomecounter: Direct Fitting to Raw Data} \label{sec:deltaM_fit}

\begin{figure}[htbp]
    \centering
    \includegraphics[width =1\columnwidth]{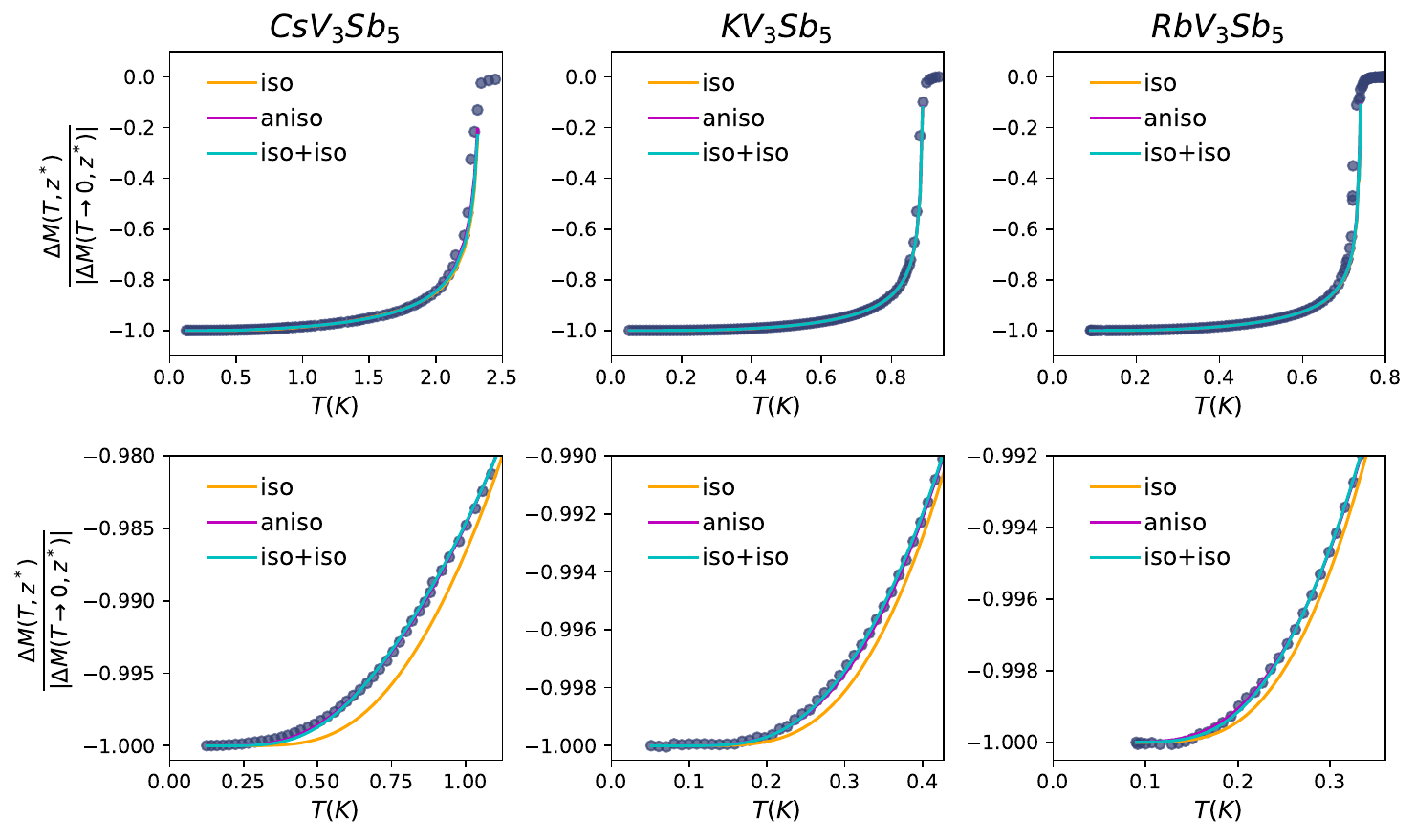}
    \caption{Fits to the raw data $\Delta M(T)$ using Eq.\ref{eqn:M_rho} (corresponding to fits for $\rho_s(T)$ in the main text) for various gap models: single isotropic gap (iso), single anisotropic gap (aniso), and two isotropic gaps (aniso). Fits over the full temperature range are shown in the top row, showing each of the gap models captures the data very well. The bottom row shows the low temperature data. All three compounds show some deviation between the single isotropic gap model, but are well described by either a single anisotropic gap or two isotropic gaps.}
    \label{fig:deltaM_fit}
\end{figure}

In Fig.\ref{fig3} in the main text, we present fits to $\rho_s(T)$ for the different gap models by fitting $\Delta M(T)$ to Eq.\ref{eqn:M_rho} with $\lambda_0/a$ being a fit parameter. The $\rho_s(T)$ data points  plotted in Fig. 3 are obtained by inverting Eq.\ref{eqn:M_rho} using the fitted value of $\lambda_0/a$. In Fig.\ref{fig:deltaM_fit} we show the raw data $\Delta M(T)$ alongside fits for the various gap models. Similar to $\rho_s(T)$, any of the gap models effectively capture the data over the full temperature range (top row of Fig.\ref{fig:deltaM_fit}). In the low temperature regime (bottom row of Fig.\ref{fig:deltaM_fit}), we observe some deviations between the data and a single isotropic gap model for all three compounds. Both the single anisotropic gap and two isotropic gaps models fit the data well, making it challenging to determine which model provides a better description of the data. The resulting fit parameter values for each compound and gap model are summarized in Table.\ref{table}. The value for $\lambda_0/a$ obtained for the various gap models are similar, and hence not very model dependent.

\newpage
\stepcounter{somecounter}
\section{Supplementary Note \thesomecounter: Resulting Fit Parameters for Fitting Procedure}

\label{sec:fit_params}
\begin{table*}[htbp]
  \centering
  \begin{tabular}{|p{2cm}||p{2.5cm}|p{2cm}|p{1cm}|p{2cm}|p{1cm}|p{0cm}|}
     \hline
     \multicolumn{7}{|c|}{Gap Fit Parameters} \\
     \hline
        & \centering $\lambda_0/a$ (x$10^{-3}$) & \centering $\Delta_{0_1}$ (meV) & \centering $\gamma$ & \centering $\Delta_{0_2}$ (meV) & \centering $\alpha$ &\\
     \hline
     \textbf{CsV$_3$Sb$_5$}  &&&&&& \\
     iso & \centering 65  & \centering  0.31 & \centering & && \\
     aniso & \centering 74 & \centering 0.53 & \centering 0.63 & && \\
     iso+iso & \centering 69 & \centering 0.23 & \centering  & \centering 0.60 & \centering 0.51& \\
     &&&&&&\\
     \textbf{KV$_3$Sb$_5$}  &&&&&& \\
     iso & \centering 41  & \centering 0.14 & \centering & && \\
     aniso & \centering 42 & \centering 0.17 & \centering 0.39 & && \\
     iso+iso & \centering 41 & \centering 0.12 & \centering  & \centering 0.24 & \centering 0.64 &\\
     &&&&&& \\
     \textbf{RbV$_3$Sb$_5$}  &&&&&& \\
     iso & \centering 40  & \centering 0.11 & \centering & && \\
     aniso & \centering 42 & \centering 0.14 & \centering 0.42 & && \\
     iso+iso & \centering 41 & \centering 0.10 & \centering  & \centering 0.26 & \centering 0.75 &\\
     \hline
  \end{tabular}
  \caption{Resulting fit parameters for each compound and different gap models. }
  \label{table}
\end{table*}

\begin{figure}
    \centering
    \includegraphics[width =1\columnwidth]{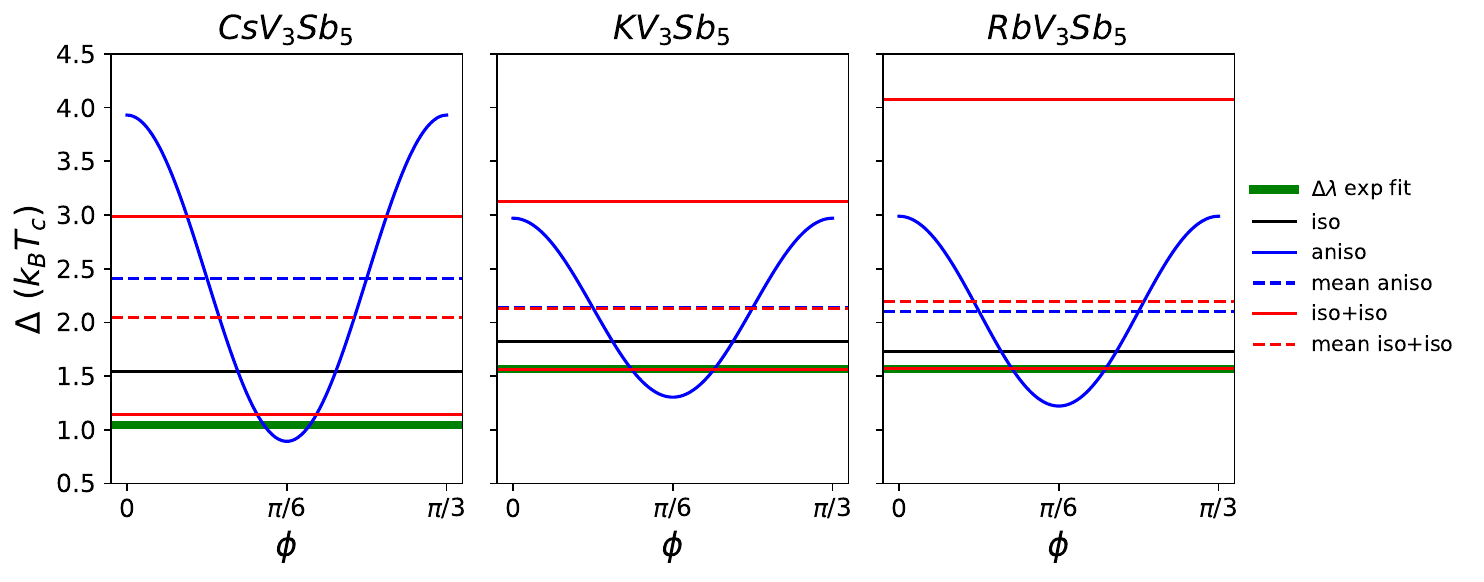}
    \caption{Comparison of the gaps for each of the three AV$_3$Sb$_5$ compounds for different fits and gap models. ($\Delta\lambda$ exp fit) fitting the low temperature range of $\Delta\lambda(T)$ up to 0.3$T/T_c$ with the exponential expression Eq.\ref{eqn:exp} (see Fig.\ref{fig:fig2}); (iso) single isotropic gap fit; (aniso) single anisotropic gap fit; (mean aniso) mean of the single anisotropic gap fit over a $\uppi$/3 period; (iso+iso) two isotropic gap fit; (mean iso+iso) weighted mean of the two isotropic gap fit with weighting factor $\alpha$. For all compounds, the gap extracted for the exponential fit of the low-temperature behavior of $\Delta\lambda(T)$ corresponds to the smaller gap in the multigap fit. The minimum of the gap in the anisotropic model is smaller than the small gap in the multigap fit to account for the low-temperature behavior with sufficient weight over the angle $\phi$. The average gap values from the anisotropic and multigap models are reasonably consistent. In KV$_3$Sb$_5$ and RbV$_3$Sb$_5$, the larger gap in the multigap model takes on a higher value, but larger gap takes on a lower weighting factor, suggesting strong correlation between the gap magnitude and the weighting ratio and limited constraint on the larger gap value. RbV$_3$Sb$_5$ and KV$_3$Sb$_5$ show similar values and trends consistent with the similarity in the measured $\Delta\lambda(T)$, while CsV$_3$Sb$_5$ behaves distinctly. For CsV$_3$Sb$_5$ more weight is effectively assigned to the larger gap reflected in a larger $\alpha$ in the multigap fit and greater anisotropy in the anisotropic model.}
    \label{fig:gap_compare}
\end{figure}

\newpage
\bibliography{bib.bib}
\makeatletter\@input{xx2.tex}\makeatother